\documentclass{article}
\usepackage{spconf,amsmath,graphicx}
\usepackage{booktabs}
\usepackage{hyperref}
\usepackage{multirow}
\usepackage{xcolor}

\title{Audio Deepfake Verification}
\name{Li Wang, Junyi Ao, Linyong Gan, Yuancheng Wang, Xueyao Zhang, Zhizheng Wu\thanks{Thanks to XYZ agency for funding.}}

\address{
\begin{tabular}{c}
School of Data Science, Shenzhen Research Institute of Big Data, \\ The Chinese University of Hong Kong, Shenzhen (CUHK-Shenzhen), China
\end{tabular}
}

\begin{document}
%
\maketitle
\begin{abstract}
With the rapid development of deepfake technology, simply making a binary judgment of true or false on audio is no longer sufficient to meet practical needs. Accurately determining the specific deepfake method has become crucial. This paper introduces the Audio Deepfake Verification (ADV) task, effectively addressing the limitations of existing deepfake source tracing methods in closed-set scenarios, aiming to achieve open-set deepfake source tracing. Meanwhile, the Audity dual-branch architecture is proposed, extracting deepfake features from two dimensions: audio structure and generation artifacts. 
Experimental results show that the dual-branch Audity architecture outperforms any single-branch configuration, and it can simultaneously achieve excellent performance in both deepfake detection and verification tasks.

\end{abstract}
\section{Introduction}
\label{sec:intro}

Artificially generated synthetic media, or \textbf{deepfakes}, have advanced rapidly over the past decade, enabling creative expression but also posing severe risks -- from cybercrime to political disinformation \cite{westerlund2019emergence}. In audio processing, while \textbf{Audio Deepfake Detection (ADD)} identifies whether an audio is synthetic via binary classification \cite{li2024audioantispoofingdetectionsurvey,yi2023audio}, the emerging task of \textbf{deepfake source tracing} (or attribution \cite{yan2024audiodeepfakeattributioninitial}, fingerprint~\cite{yan2022initial}, algorithm recognition~\cite{yi2023add}) aims to identify the specific generative model or tool used.
This goes beyond detection, providing critical evidence for digital forensics and accountability.

Most existing audio deepfake tracers operate in \textbf{closed-set scenarios}, only recognizing predefined algorithms and treating unknowns as a generic ``unknown" class \cite{Klein2024}. Two common strategies are threshold-based similarity scoring \cite{yan2024audiodeepfakeattributioninitial} and anomaly detection \cite{xie24_interspeech}; however, both approaches are unable to attribute audio generated by previously unseen methods. 

In particular, Wang et al. \cite{wang2025generalize} proposed an audio deepfake attribution enhancement strategy (ADAE) that decouples speaker identity from generation-style features via a speaker encoder and information disentangle block. While this approach effectively mitigates speaker interference in feature space, it primarily focuses on removing time-invariant speaker information to isolate algorithm-specific fingerprints (e.g., via differential rectification of speaker-parallel components). However, this strict disentanglement may inadvertently discard subtle speaker-deepfake interaction cues, including the transformation patterns of specific vocal characteristics induced by different synthesis algorithms. Such cues are critical for cross-speaker generalization in open-set scenarios, where the model must recognize unseen algorithms across diverse speaker identities.

Inspired by paradigms in speaker verification, we introduce the \textbf{Audio Deepfake Verification (ADV)} task. ADV aims to determine whether two audio samples were generated using the same deep fake method, \textbf{ enabling open set source tracking without restricting to predefined algorithms}.

Effective feature extraction is paramount for Audio Deepfake Verification (ADV), a task that extends beyond the binary authenticity classification of Audio Deepfake Detection (ADD). ADV necessitates the identification of discriminative features that not only confirm the artificial origin of an audio sample but also, crucially, reveal the characteristic signatures of the specific deepfake generation methodology employed.

Currently, state-of-the-art audio deepfake feature extraction often leverages self-supervised pre-trained models such as Wav2Vec \cite{baevski2020wav2vec} and WavLM \cite{chen2022wavlm}. These models, typically trained through masked prediction objectives, are highly effective at learning robust representations of natural speech structures, including phoneme sequences, prosodic patterns, and linguistic temporal dependencies.

However, for the task of ADV, it is not sufficient to rely solely on speech structural features. In addition to capturing the natural speech structure, it is also crucial to extract generation-specific artifacts, such as discontinuous high-frequency energy distributions, anomalous patterns in silent segments, breath sounds, pauses, and spectral distortions introduced during synthesis. These artifacts can provide important cues for distinguishing between different deepfake generation methods.

Therefore, in the context of ADV, it is essential to jointly consider both the structural features of natural speech and the generation-specific artifacts. To this end, we propose the dual-branch Audity architecture, which is specifically designed to extract and integrate these two complementary types of features.

Experimental results demonstrate the effectiveness of the proposed Audity architecture. Notably, Audity exhibits strong verification capabilities not only on commercial audio deepfake systems but also on state-of-the-art deepfake generation models. It is worth highlighting that Audity achieves high performance on both verification and detection tasks simultaneously, showcasing its versatility and robustness in practical audio deepfake scenarios.


    

\section{Audio Deepfake Verification}

\subsection{Definition}

Currently, existing methods for deepfake source tracing are predominantly designed for closed-set scenarios or classify unseen deepfake methods as ``unknown".
To recognize new deepfake methods, retraining the models becomes necessary. Inspired by speaker verification paradigms, this paper introduces the Audio Deepfake Verification (ADV) task, aiming to achieve open-set deepfake source tracing.

\begin{figure}[h]
    \centering
    \includegraphics[width=1\linewidth]{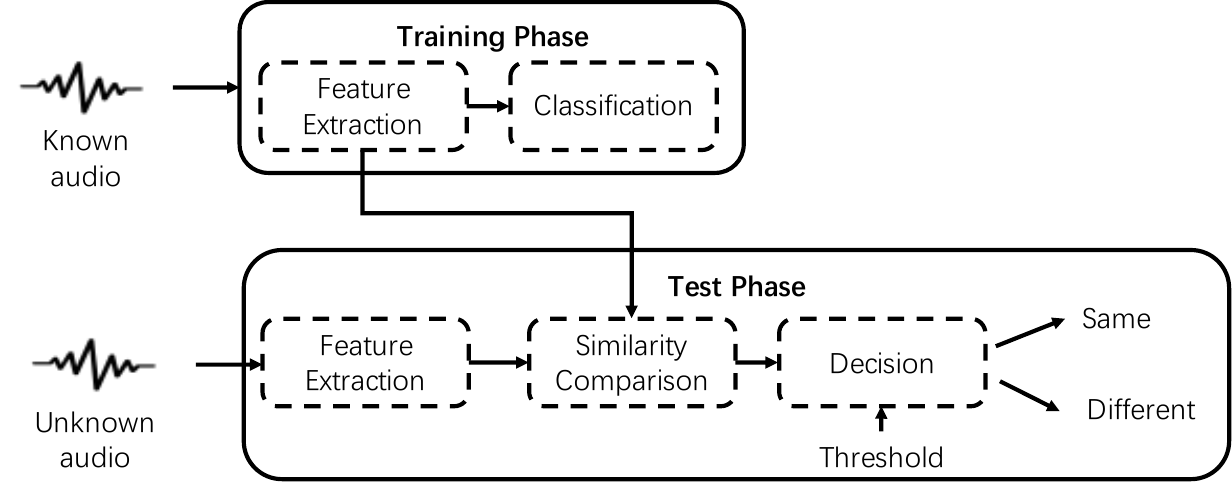}
    \caption{Block diagram of the Audio Deepfake Verification (ADV) system.}
    \label{fig:adv_dig}
\end{figure}

The basic block diagram of the ADV system is shown in Figure~\ref{fig:adv_dig}. In the training phase, data from known deepfake methods are utilized for closed-set training, with the objective of extracting deepfake features. During the testing phase, an unknown sample is compared with the features of known samples. If the similarity score exceeds the threshold, it is considered to be generated by the same deepfake method; otherwise, it is deemed to be generated by a different method.

\subsection{Audity: A Dual-Branch Network for Audio Deepfake Verification}
\begin{figure}[h]
    \centering
    \includegraphics[width=1\linewidth]{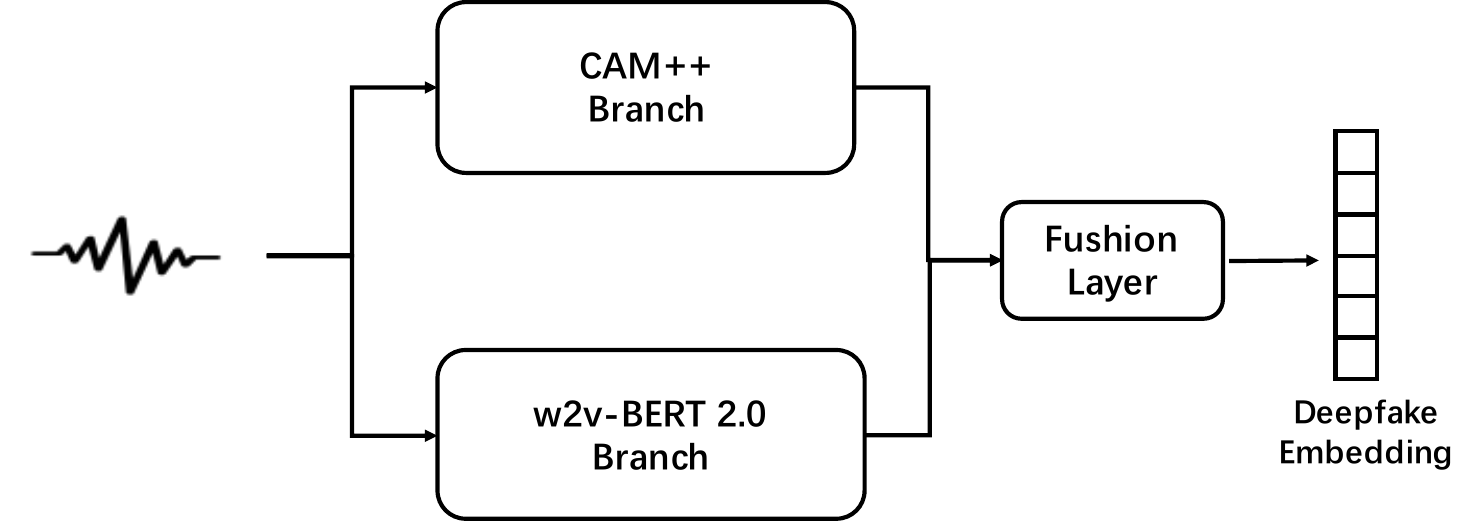}
    \caption{The overall architecture of Audity.}
    \label{fig:ASNet}
\end{figure}

To address the complementary modeling of speech structure and generation artifacts, we propose \textbf{Audity}, a dual-branch architecture that integrates pre-trained speech structural representations with generation artifact modeling. As illustrated in Figure~\ref{fig:ASNet}, the network consists of two parallel pathways: an \textbf{Audio Structural Branch} and a \textbf{Generation Artifacts Branch}, which are fused to produce discriminative deepfake embeddings.

\subsubsection{Audio Structural Branch}

Leveraging the strengths of self-supervised pre-training, this branch employs w2v-BERT 2.0 ~\cite{barrault2023seamless}\footnote{\href{https://huggingface.co/facebook/w2v-bert-2.0}{https://huggingface.co/facebook/w2v-bert-2.0}} to extract high-level structural representations of speech content. Unlike traditional ``semantic" modeling, which focuses on linguistic meaning, our design emphasizes capturing temporal dependencies of language units - including phoneme sequences, prosodic contours, and syllabic rhythms - that characterize natural speech consistency. Input as a mel-spectrogram, the backbone processes sequential features through masked prediction objectives, learning to reconstruct intact speech structures from corrupted inputs. This enables the branch to encode normative patterns of human speech production and detect structural anomalies introduced by deepfake algorithms (e.g., unnatural prosodic breaks or phoneme misalignments).

\subsubsection{Generation Artifacts Branch} 

To model artifacts specific to synthetic audio generation, this branch is designed to flexibly accommodate various state-of-the-art acoustic modeling architectures. By taking spectrograms as input, the branch preserves comprehensive raw audio information, enabling the extraction of generation-specific artifacts without reliance on handcrafted features. This approach allows the model to capture technical fingerprints associated with different deepfake generation methods.

\subsubsection{Feature Fusion and Embedding Generation}

The outputs of the two branches are concatenated and passed through a fusion layer, which integrates the audio structural features and generation artifact features. The resulting joint representation captures both the intrinsic patterns of natural speech and the distinctive artifacts introduced by synthetic generation. This comprehensive feature integration enables Audity to support the ADV task, facilitating not only the detection of deepfake audio but also the accurate attribution of its generative source.

\section{Experimental Setups}
\subsection{Dataset}

\subsubsection{Training Dataset}
In this study, we design and employ two distinct training settings to comprehensively evaluate the effectiveness of Audity.

First, to ensure rigorous benchmarking and comparability, we adopt the official protocol of MLAAD SourceTrace~\cite{UsingMLAADforSourceTracing}\footnote{\href{https://deepfake-total.com/sourcetracing}{https://deepfake-total.com/sourcetracing}}, where the training set consists of samples generated by 24 different audio deepfake methods.

Second, only through training with diverse deepfake methods can the model effectively extract the key features of various deepfake techniques. Therefore, this study has comprehensively collected 10 open-source datasets, and the detailed statistical information is presented in Table \ref{tab:train_set}. All datasets are partitioned according to their official splits whenever available. For datasets without official training, validation, and test splits, the entire dataset is utilized as training data.

\begin{table}[ht]
\centering
\small
\caption{Training dataset}
\label{tab:train_set}
\begin{tabular}{lrrc}
  \toprule
  \textbf{Dataset} & \textbf{\#Utterance} & \textbf{Hours} & \textbf{\#Types} \\
  \midrule
   ASVspoof2019-LA~\cite{nautsch2021asvspoof}& 25,380 & 24.15  & 6 \\
   CodecFake(UCAS)~\cite{xie2025codecfake} & 740,747 & 706.43 & 6 \\
   CodecFake+~\cite{chen2025codecfakelargescaleneuralaudio}& 1,417,845 & 1,341.30 & 31 \\
   DFADD~\cite{du2024dfadd} & 372,325 & 373.97 & 5 \\
   GigaSpeech(M)~\cite{chen2021gigaspeech}& 885,397 & 973.44 & 0 \\
   LibriSeVoc~\cite{sun2023ai}& 55,440 & 145.02 & 6 \\
   MLAAD~\cite{muller2024mlaad}& 151,446 & 371.21 & 49 \\
   SpoofCeleb~\cite{jung2025spoofceleb}& 2,540,421 & 1,982.23 & 10 \\
   Wavefake~\cite{frank2021wavefake}& 110,870 & 197.91 & 9 \\
   ADD2023 Track3~\cite{yi2024add}& 22,400 & 34.93 & 6 \\
 \cmidrule(lr){1-4}
   \textbf{Total} & \textbf{6,322,271} & \textbf{6,150.60} & \textbf{128} \\
  \bottomrule
\end{tabular}
\end{table}

The validation dataset is constructed based on the official validation splits provided by each source dataset. The detailed statistical information for the validation set is summarized in Table~\ref{tab:valid_set}.

\begin{table}[ht]
\centering
\small
\caption{Validate dataset}
\label{tab:valid_set}
\begin{tabular}{lrrc}
  \toprule
  \textbf{Dataset} & \textbf{\#Utterance} & \textbf{Hours} & \textbf{\#Types} \\
  \midrule
  ASVspoof2019-LA~\cite{nautsch2021asvspoof} & 24,844 & 24.00 & 6 \\
  CodecFake(UCAS)~\cite{xie2025codecfake} & 92,596 & 88.66 & 6 \\
  CodecFake+~\cite{chen2025codecfakelargescaleneuralaudio} & 48,510 & 46.13 & 23 \\
  DFADD~\cite{du2024dfadd} & 6,675 & 6.78 & 5 \\
  GigaSpeech(M)~\cite{chen2021gigaspeech} & 5,715 & 11.37 & 0 \\
  LibriSeVoc~\cite{sun2023ai} & 18,480 & 49.03 & 6 \\
  SpoofCeleb~\cite{jung2025spoofceleb} & 55,741 & 40.92 & 6 \\
  ADD2023 Track3~\cite{yi2024add} & 8,400 & 13.90 & 6 \\
    \cmidrule(lr){1-4}
   \textbf{Total} & \textbf{260,961} & \textbf{280.78} & \textbf{58} \\
  \bottomrule
\end{tabular}
\end{table}

Regarding the annotation of deepfake methods, there is currently no consensus in the academic community on the factors associated with deepfake fingerprints. It is generally believed that they may be closely related to model architectures, optimization algorithms, and training datasets. In view of this, this study has conducted a rigorous examination of the deepfake methods in each dataset. Identical annotations are only assigned to models that are completely the same, thus maximizing the retention of rich meta-information and laying a foundation for subsequent research.

\subsubsection{Test Dataset}




We conducted evaluations using five different test sets, which are shown in Table~\ref{tab:test_set}. Especially, \textbf{Demo Page datasets} are collected from various demo pages and include samples generated by CosyVoice~\cite{du2024cosyvoice}, CosyVoice2\cite{du2024cosyvoice2}, E2TTS\cite{eskimez2024e2}, FireRedTTS\cite{guo2024fireredtts}, Llasa\cite{ye2025llasa}, MaskGCT\cite{wang2024maskgct}, SeedTTS\cite{anastassiou2024seed}, and SparkTTS\cite{wang2025spark}, while \textbf{Commercial TTS datasets} consist of audio generated by commercial Text-to-Speech (TTS) systems, including the latest gpt-4o-mini-tts from OpenAI and audio produced by Microsoft Azure TTS.

\begin{table}[ht]
\centering
\small
\caption{Test dataset}
\label{tab:test_set}
\begin{tabular}{lrrc}
  \toprule
  \textbf{Dataset} & \textbf{\#Utterance} & \textbf{Hours} & \textbf{\#Types} \\
  \midrule
  ASVspoof2019-LA~\cite{nautsch2021asvspoof} & 71,237 & 61.49 & 13 \\
  SpoofCeleb~\cite{jung2025spoofceleb} & 91,130 & 66.08 & 9 \\
  ADD2023 Track3~\cite{yi2024add} & 79,490 & 119.43  & 7 \\
  Demo Page dataset & 604 & 2.04  & 8 \\
  Commercial TTS dataset & 200 & 0.25 & 2 \\
   \bottomrule
\end{tabular}
\end{table}

\subsection{Experimental Setups}

\textbf{Construction of test sample pairs:} For the samples in each test set, samples of the same type and different types are randomly selected to form verification sample pairs.

\textbf{Validation set and model selection:} The validation set is shown in Table~\ref{tab:valid_set} and all come from the official division. Sample pairs are constructed using the same method as that for the test set. The model with the lowest Equal Error Rate (EER) on the validation set is selected as the optimal model. The threshold corresponds to the EER of the validation set.

\textbf{Sample type distribution balance:} At the start of each epoch, randomly sample 1000 samples per type to ensure the balance of every type.

\textbf{Implementation details of the generation artifacts branch}: We implemented the generation artifacts branch using several alternative architectures, including CAM++ \cite{wang2023cam++}, ECAPA-TDNN \cite{desplanques2020ecapa}, and ResNet293\cite{he2016deep}. All implementations were developed based on the WeSpeaker toolkit\cite{wang2024advancing,wang2023wespeaker}, ensuring consistency and reproducibility across different model variants.

\subsection{Metric}
The test metrics consist of six items: Equal Error Rate (EER), Accuracy (Acc), False Acceptance Rate (FAR), False Rejection Rate (FRR), F1 score, and Area Under the Receiver Operating Characteristic curve (AUROC). Among them, FAR represents the sample pair with ``different" label that are misidentified as ``same".

In addition, to provide a reference for practical applications, the FRR when the FAR is 1\% and the FAR when the FRR is 1\% are also reported.

\section{Experimental Results}

\subsection{Performance of Different Audity Architectures}

\begin{table}[h] 
\centering
\small
\caption{Results of different Audity branch configurations on the MLAAD SourceTrace dataset. 
GAB: Generation Artifacts Branch; ASB: Audio Structural Branch.}
\label{tab:audity_arch}
\begin{tabular}{llcc}
  \toprule
  \textbf{GAB} & \textbf{ASB} & \textbf{Acc (\%)} & \textbf{EER (\%)} \\
  \midrule
  ECAPA-TDNN  & w2v-BERT 2.0 & 89.28 & 10.49 \\
  ResNet239   & w2v-BERT 2.0 & 79.80 & 21.05 \\
  CAM++       & w2v-BERT 2.0 & \textbf{89.87} & \textbf{10.02} \\
  -           & w2v-BERT 2.0 & 86.69 & 13.29 \\
  CAM++       & -            & 68.77 & 31.35 \\
  
  \bottomrule
\end{tabular}
\end{table}

\textbf{Integrating different Generation Artifacts Branches with the Audio Structural Branch reveals that CAM++ achieves the best overall performance.} Table~\ref{tab:audity_arch} presents the effectiveness of the Audity architecture with different branch configurations on the MLAAD SourceTrace dataset. The first three rows compare the impact of various Generation Artifacts Branch (GAB) implementations, including ECAPA-TDNN, ResNet239, and CAM++, each combined with the Audio Structural Branch (ASB, implemented as w2v-BERT 2.0). Among these configurations, the combination of CAM++ and w2v-BERT 2.0 achieves the best performance, with an accuracy of 89.87\% and an EER of 10.02\%.

\textbf{The dual-branch architecture outperforms either single branch alone, highlighting the complementarity of structural and artifact features.} The last two rows provide further insights by evaluating single-branch configurations. Using only the Audio Structural Branch (w2v-BERT 2.0) without a Generation Artifacts Branch yields an accuracy of 86.69\% and an EER of 13.29\%. In contrast, employing only the Generation Artifacts Branch (CAM++) without the Audio Structural Branch results in a significant performance drop, with an accuracy of 68.77\% and an EER of 31.35\%. These results demonstrate that the dual-branch architecture, which integrates both structural and generation artifact features, substantially outperforms either branch alone.

\subsection{Performance of Audity on Multiple Test Datasets}

\begin{table*}[h] 
\centering
\small
\caption{
Audio Deepfake Verification Results on Different Test Datasets using Audity with CAM++ as the Generation Artifacts Branch. 
\#E denotes the number of enrollment samples (i.e., samples with known fake types), and \#V denotes the number of verification samples (i.e., test samples).
}
\label{tab:result}
\begin{tabular}{lrrrrrrrrrrr}
\toprule
\textbf{Test Dataset} & \textbf{\#E} & \textbf{\#V} 
& \textbf{Acc} & \textbf{FAR} & \textbf{FRR} & \textbf{EER} 
& \textbf{F1} & \textbf{AUROC} & \textbf{FRR@FAR1\%} & \textbf{FAR@FRR1\%} \\
\midrule
\multirow{1}{*}{SpoofCeleb}         
    & 1  & 1  & 87.3\% & 19.9\% & 5.6\%  & 13.7\% & 0.88 & 0.94 & 76.0\% & 26.6\% \\
\midrule
\multirow{1}{*}{ASVspoof2019}       
    & 1  & 1  & 88.6\% & 15.7\% & 7.1\%  & 11.6\% & 0.89 & 0.95 & 67.3\% & 44.9\% \\
\midrule
\multirow{1}{*}{Commericial System} 
    & 1  & 1  & 73.5\% & 9.5\%  & 43.5\% & 24.0\% & 0.68 & 0.83 & 68.5\% & 92.0\% \\
\midrule
\multirow{3}{*}{ADD2023 Track3}     
    & 1  & 1  & 76.4\% & 14.2\% & 33.1\% & 25.0\% & 0.74 & 0.83 & 65.3\% & 95.1\% \\
    & 5  & 1  & 85.2\% & 14.7\% & 14.8\% & 14.8\% & 0.85 & 0.92 & 52.0\% & 75.6\% \\
    & 5  & 5  & 89.1\% & 21.4\% & 0.5\%  & 4.2\%  & 0.90 & 0.99 & 31.2\% & 8.6\%  \\
\midrule
\multirow{3}{*}{Demo Pages}         
    & 1  & 1  & 62.8\% & 17.6\% & 57.0\% & 37.2\% & 0.54 & 0.68 & 91.1\% & 97.4\% \\
    & 5  & 1  & 67.8\% & 47.0\% & 17.4\% & 30.8\% & 0.72 & 0.76 & 89.1\% & 88.3\% \\
    & 5  & 5  & 57.9\% & 84.1\% & 0.2\%  & 17.0\% & 0.70 & 0.90 & 88.7\% & 54.8\% \\
\bottomrule
\end{tabular}
\end{table*}

\textbf{More advanced deepfake generation methods may be more difficult to distinguish due to their increased realism.}
The results of the Audity model trained on the datasets summarized in Table~\ref{tab:train_set} are presented in Table~\ref{tab:result}. As shown, there is a considerable variation in performance across different test datasets. Notably, the ADD2023 Track3, Commercial System, and Demo Pages datasets present the greatest challenges. This can be attributed to the fact that ADD2023 Track3 contains a portion of commercial data, while the Demo Pages dataset consists exclusively of samples generated by the latest deepfake models. The presence of these advanced and diverse generation methods increases the difficulty of distinguishing between different types of audio deepfake methods.

\textbf{Using the average of embeddings from a larger number of enrollment samples to represent each deepfake method leads to improved performance.} We conducted experiments on ADD2023 Track3 and Demo Pages by increasing both the number of enrollment and verification samples to 5, and using the mean of their embeddings as the representation for each method. As shown in Table~\ref{tab:result}, when both the enrollment and verification sample numbers are set to 5, the EER on ADD2023 Track3 drops significantly to 4.2\%, and on Demo Pages to 17.0\%. This demonstrates a substantial improvement compared to the single-sample setting. This improvement is reasonable, as incorporating more samples allows the model to capture more representative and robust features of each deepfake method.

\subsection{Visualization of Deepfake Verification Embeddings}

\begin{figure*}[h]
    \centering
    \includegraphics[width=1\linewidth]{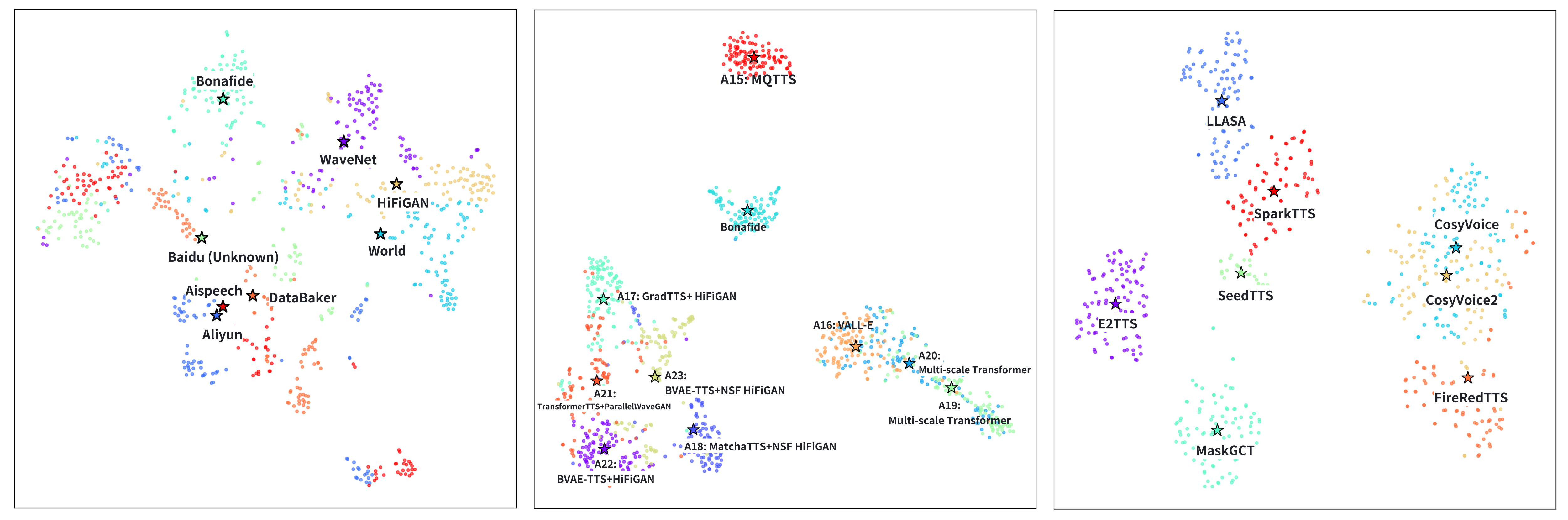}
    \caption{
    Visualization of deepfake audio embeddings using t-SNE. Each color represents a different deepfake generation method, and the star markers indicate the cluster centers for each method. The left panel corresponds to ADD2023 Track3, the middle panel to SpoofCeleb, and the right panel to our self-collected Demo Page dataset. 
    For SpoofCeleb (middle), the authors use ``Acoustic model" and ``Waveform model" to denote different deepfake methods \cite{jung2025spoofceleb}.
    }
    \label{fig:visu_emb}
\end{figure*}

Figure~\ref{fig:visu_emb} presents the visualization results for three datasets: the left panel corresponds to ADD2023 Track3, the middle panel to SpoofCeleb, and the right panel to the Demo Page dataset.

\textbf{Models with similar architectures may exhibit similar deepfake feature distributions.} For example, in the SpoofCeleb results (middle panel), both A19 and A20 are based on the Multi-scale Transformer architecture, while the waveform models for A18 and A23 are both NSF HiFiGAN. Similarly, in the Demo Page results (right panel), CosyVoice and CosyVoice2 display highly similar feature distributions. However, it is important to note that research on the factors associated with deepfake features is still in its early stages. The current consensus in the research community is that model architecture plays a significant role, but other factors related to data distribution, such as training data and optimization algorithms, have received relatively little attention.

\textbf{Commercial models are difficult to distinguish from each other.} As shown in the visualization of ADD 2023 (left panel), the feature distributions of commercial systems such as Baidu and Aispeech are highly concentrated and challenging to separate. This may be attributed to the fact that commercial systems often adopt the most robust solutions in the industry, leading to similarities in model architectures and optimization strategies. Additionally, these systems may collect their own large-scale datasets, and as the amount of data increases, the data distributions may also become more similar.

\subsection{Can the Audio Deepfake Verification Model be directly used for Audio Deepfake Detection?}
As shown in Figure~\ref{fig:visu_emb}, Bonafide samples are well separated from various deepfake methods. Motivated by this observation, we further explored using the posterior probability of Bonafide as the detection score for deepfake detection (i.e., binary classification). All test sets follow the official evaluation protocols. The results are summarized in Table~\ref{tab: detect}.

\textbf{Learning to distinguish features of different deepfake methods enhances the model's understanding of deepfakes.} Unlike simple binary classification, which treats all deepfakes as a single ``spoof" class, differentiating between various deepfake methods allows the model to capture the unique ``fingerprints" of each technique. This richer understanding enables the model to identify more subtle and discriminative patterns, resulting in a more comprehensive representation of deepfake characteristics and improved performance in both verification and detection tasks across diverse scenarios.

\begin{table}[h]
\centering
\caption{Audio Deepfake Detection Results compared with previous state-of-the-art (SOTA) methods.}
\label{tab: detect}
\begin{tabular}{lcc}
\toprule
\textbf{Test Dataset} & \textbf{EER (Ours) (\%)} & \textbf{EER (SOTA) (\%)} \\
\midrule
In-the-Wild         & 7.14  & 7.68 \\
SpoofCeleb          & 0.07  & 1.12 \\
ASVspoof2019        & 0.03  & 0.22 \\
\bottomrule
\end{tabular}
\end{table}

\textbf{Data diversity is more important than quantity.} As shown in Table~\ref{tab:train_set}, we utilized GigaSpeech~\cite{chen2021gigaspeech}, a highly diverse dataset composed entirely of bonafide samples. This diversity helps the model better understand the natural variability of human speech, including different accents, speaking styles, and recording conditions. Exposure to such a wide range of real-world speech enables the model to establish a robust baseline for authentic audio, making it more effective at detecting subtle deviations caused by deepfake manipulations and thereby improving performance in both verification and detection tasks.

\section{Conclusions}

In this paper, we introduced the Audio Deepfake Verification (ADV) task to address the limitations of existing closed-set deepfake source tracing methods, enabling open-set source attribution by determining whether two audio samples are generated by the same deepfake method. We proposed Audity, a dual-branch architecture comprising an Audio Structural Branch, which encodes the temporal consistency of natural speech, and a Generation Artifacts Branch, which captures unique artifacts specific to synthetic audio. Experiments across diverse datasets demonstrated the effectiveness of Audity in distinguishing between different deepfake generation methods, although more advanced methods remain challenging. Remarkably, our model also achieved excellent performance in Audio Deepfake Detection on several datasets, suggesting that Audity can simultaneously perform both audio deepfake detection and verification tasks.



\bibliographystyle{IEEEbib}
\bibliography{strings,refs}

\begin{thebibliography}{10}

\bibitem{westerlund2019emergence}
Mika Westerlund,
\newblock ``The emergence of deepfake technology: A review,''
\newblock {\em Technology innovation management review}, vol. 9, no. 11, 2019.

\bibitem{li2024audioantispoofingdetectionsurvey}
Menglu Li, Yasaman Ahmadiadli, and Xiao-Ping Zhang,
\newblock ``Audio anti-spoofing detection: A survey,'' 2024.

\bibitem{yi2023audio}
Jiangyan Yi, Chenglong Wang, Jianhua Tao, Xiaohui Zhang, Chu~Yuan Zhang, and Yan Zhao,
\newblock ``Audio deepfake detection: A survey,''
\newblock {\em arXiv preprint arXiv:2308.14970}, 2023.

\bibitem{yan2024audiodeepfakeattributioninitial}
Xinrui Yan, Jiangyan Yi, Jianhua Tao, and Jie Chen,
\newblock ``Audio deepfake attribution: An initial dataset and investigation,'' 2024.

\bibitem{yan2022initial}
Xinrui Yan, Jiangyan Yi, Jianhua Tao, Chenglong Wang, Haoxin Ma, Tao Wang, Shiming Wang, and Ruibo Fu,
\newblock ``An initial investigation for detecting vocoder fingerprints of fake audio,''
\newblock in {\em Proceedings of the 1st International Workshop on Deepfake Detection for Audio Multimedia}, 2022, pp. 61--68.

\bibitem{yi2023add}
Jiangyan Yi, Jianhua Tao, Ruibo Fu, Xinrui Yan, Chenglong Wang, Tao Wang, Chu~Yuan Zhang, Xiaohui Zhang, Yan Zhao, Yong Ren, et~al.,
\newblock ``Add 2023: the second audio deepfake detection challenge,''
\newblock {\em arXiv preprint arXiv:2305.13774}, 2023.

\bibitem{Klein2024}
Nicholas Klein, Tianxiang Chen, Hemlata Tak, Ricardo Casal, and Elie Khoury,
\newblock ``Source tracing of audio deepfake systems,''
\newblock in {\em Interspeech 2024}. Sept. 2024, p. 1100–1104, ISCA.

\bibitem{xie24_interspeech}
Yuankun Xie, Ruibo Fu, Zhengqi Wen, Zhiyong Wang, Xiaopeng Wang, Haonnan Cheng, Long Ye, and Jianhua Tao,
\newblock ``Generalized source tracing: Detecting novel audio deepfake algorithm with real emphasis and fake dispersion strategy,''
\newblock in {\em Interspeech 2024}, 2024, pp. 4833--4837.

\bibitem{wang2025generalize}
Zhigang Wang, Dengpan Ye, Jingyang Li, and Jiacheng Deng,
\newblock ``Generalize audio deepfake algorithm recognition via attribution enhancement,''
\newblock in {\em ICASSP 2025-2025 IEEE International Conference on Acoustics, Speech and Signal Processing (ICASSP)}. IEEE, 2025, pp. 1--5.

\bibitem{baevski2020wav2vec}
Alexei Baevski, Yuhao Zhou, Abdelrahman Mohamed, and Michael Auli,
\newblock ``wav2vec 2.0: A framework for self-supervised learning of speech representations,''
\newblock {\em Advances in neural information processing systems}, vol. 33, pp. 12449--12460, 2020.

\bibitem{chen2022wavlm}
Sanyuan Chen, Chengyi Wang, Zhengyang Chen, Yu~Wu, Shujie Liu, Zhuo Chen, Jinyu Li, Naoyuki Kanda, Takuya Yoshioka, Xiong Xiao, et~al.,
\newblock ``Wavlm: Large-scale self-supervised pre-training for full stack speech processing,''
\newblock {\em IEEE Journal of Selected Topics in Signal Processing}, vol. 16, no. 6, pp. 1505--1518, 2022.

\bibitem{barrault2023seamless}
Lo{\"\i}c Barrault, Yu-An Chung, Mariano~Coria Meglioli, David Dale, Ning Dong, Mark Duppenthaler, Paul-Ambroise Duquenne, Brian Ellis, Hady Elsahar, Justin Haaheim, et~al.,
\newblock ``Seamless: Multilingual expressive and streaming speech translation,''
\newblock {\em arXiv preprint arXiv:2312.05187}, 2023.

\bibitem{UsingMLAADforSourceTracing}
Nicolas M{\"u}ller,
\newblock ``Using mlaad for source tracing of audio deepfakes,'' \url{https://deepfake-total.com/sourcetracing}, 11 2024.

\bibitem{nautsch2021asvspoof}
Andreas Nautsch, Xin Wang, Nicholas Evans, Tomi~H Kinnunen, Ville Vestman, Massimiliano Todisco, H{\'e}ctor Delgado, Md~Sahidullah, Junichi Yamagishi, and Kong~Aik Lee,
\newblock ``Asvspoof 2019: spoofing countermeasures for the detection of synthesized, converted and replayed speech,''
\newblock {\em IEEE Transactions on Biometrics, Behavior, and Identity Science}, vol. 3, no. 2, pp. 252--265, 2021.

\bibitem{xie2025codecfake}
Yuankun Xie, Yi~Lu, Ruibo Fu, Zhengqi Wen, Zhiyong Wang, Jianhua Tao, Xin Qi, Xiaopeng Wang, Yukun Liu, Haonan Cheng, et~al.,
\newblock ``The codecfake dataset and countermeasures for the universally detection of deepfake audio,''
\newblock {\em IEEE Transactions on Audio, Speech and Language Processing}, 2025.

\bibitem{chen2025codecfakelargescaleneuralaudio}
Xuanjun Chen, Jiawei Du, Haibin Wu, Lin Zhang, I-Ming Lin, I-Hsiang Chiu, Wenze Ren, Yuan Tseng, Yu~Tsao, Jyh-Shing~Roger Jang, and Hung yi~Lee,
\newblock ``Codecfake+: A large-scale neural audio codec-based deepfake speech dataset,'' 2025.

\bibitem{du2024dfadd}
Jiawei Du, I-Ming Lin, I-Hsiang Chiu, Xuanjun Chen, Haibin Wu, Wenze Ren, Yu~Tsao, Hung-yi Lee, and Jyh-Shing~Roger Jang,
\newblock ``Dfadd: The diffusion and flow-matching based audio deepfake dataset,''
\newblock in {\em 2024 IEEE Spoken Language Technology Workshop (SLT)}. IEEE, 2024, pp. 921--928.

\bibitem{chen2021gigaspeech}
Guoguo Chen, Shuzhou Chai, Guanbo Wang, Jiayu Du, Wei-Qiang Zhang, Chao Weng, Dan Su, Daniel Povey, Jan Trmal, Junbo Zhang, et~al.,
\newblock ``Gigaspeech: An evolving, multi-domain asr corpus with 10,000 hours of transcribed audio,''
\newblock {\em arXiv preprint arXiv:2106.06909}, 2021.

\bibitem{sun2023ai}
Chengzhe Sun, Shan Jia, Shuwei Hou, and Siwei Lyu,
\newblock ``Ai-synthesized voice detection using neural vocoder artifacts,''
\newblock in {\em Proceedings of the IEEE/CVF Conference on Computer Vision and Pattern Recognition}, 2023, pp. 904--912.

\bibitem{muller2024mlaad}
Nicolas~M M{\"u}ller, Piotr Kawa, Wei~Herng Choong, Edresson Casanova, Eren G{\"o}lge, Thorsten M{\"u}ller, Piotr Syga, Philip Sperl, and Konstantin B{\"o}ttinger,
\newblock ``Mlaad: The multi-language audio anti-spoofing dataset,''
\newblock {\em International Joint Conference on Neural Networks (IJCNN)}, 2024.

\bibitem{jung2025spoofceleb}
Jee-weon Jung, Yihan Wu, Xin Wang, Ji-Hoon Kim, Soumi Maiti, Yuta Matsunaga, Hye-jin Shim, Jinchuan Tian, Nicholas Evans, Joon~Son Chung, et~al.,
\newblock ``Spoofceleb: Speech deepfake detection and sasv in the wild,''
\newblock {\em IEEE Open Journal of Signal Processing}, 2025.

\bibitem{frank2021wavefake}
Joel Frank and Lea Sch{\"o}nherr,
\newblock ``Wavefake: A data set to facilitate audio deepfake detection,''
\newblock {\em arXiv preprint arXiv:2111.02813}, 2021.

\bibitem{yi2024add}
Jiangyan Yi, Chu~Yuan Zhang, Jianhua Tao, Chenglong Wang, Xinrui Yan, Yong Ren, Hao Gu, and Junzuo Zhou,
\newblock ``Add 2023: Towards audio deepfake detection and analysis in the wild,''
\newblock {\em arXiv preprint arXiv:2408.04967}, 2024.

\bibitem{du2024cosyvoice}
Zhihao Du, Qian Chen, Shiliang Zhang, Kai Hu, Heng Lu, Yexin Yang, Hangrui Hu, Siqi Zheng, Yue Gu, Ziyang Ma, et~al.,
\newblock ``Cosyvoice: A scalable multilingual zero-shot text-to-speech synthesizer based on supervised semantic tokens,''
\newblock {\em arXiv preprint arXiv:2407.05407}, 2024.

\bibitem{du2024cosyvoice2}
Zhihao Du, Yuxuan Wang, Qian Chen, Xian Shi, Xiang Lv, Tianyu Zhao, Zhifu Gao, Yexin Yang, Changfeng Gao, Hui Wang, et~al.,
\newblock ``Cosyvoice 2: Scalable streaming speech synthesis with large language models,''
\newblock {\em arXiv preprint arXiv:2412.10117}, 2024.

\bibitem{eskimez2024e2}
Sefik~Emre Eskimez, Xiaofei Wang, Manthan Thakker, Canrun Li, Chung-Hsien Tsai, Zhen Xiao, Hemin Yang, Zirun Zhu, Min Tang, Xu~Tan, et~al.,
\newblock ``E2 tts: Embarrassingly easy fully non-autoregressive zero-shot tts,''
\newblock in {\em 2024 IEEE Spoken Language Technology Workshop (SLT)}. IEEE, 2024, pp. 682--689.

\bibitem{guo2024fireredtts}
Hao-Han Guo, Yao Hu, Kun Liu, Fei-Yu Shen, Xu~Tang, Yi-Chen Wu, Feng-Long Xie, Kun Xie, and Kai-Tuo Xu,
\newblock ``Fireredtts: A foundation text-to-speech framework for industry-level generative speech applications,''
\newblock {\em arXiv preprint arXiv:2409.03283}, 2024.

\bibitem{ye2025llasa}
Zhen Ye, Xinfa Zhu, Chi-Min Chan, Xinsheng Wang, Xu~Tan, Jiahe Lei, Yi~Peng, Haohe Liu, Yizhu Jin, Zheqi DAI, et~al.,
\newblock ``Llasa: Scaling train-time and inference-time compute for llama-based speech synthesis,''
\newblock {\em arXiv preprint arXiv:2502.04128}, 2025.

\bibitem{wang2024maskgct}
Yuancheng Wang, Haoyue Zhan, Liwei Liu, Ruihong Zeng, Haotian Guo, Jiachen Zheng, Qiang Zhang, Xueyao Zhang, Shunsi Zhang, and Zhizheng Wu,
\newblock ``Maskgct: Zero-shot text-to-speech with masked generative codec transformer,''
\newblock {\em arXiv preprint arXiv:2409.00750}, 2024.

\bibitem{anastassiou2024seed}
Philip Anastassiou, Jiawei Chen, Jitong Chen, Yuanzhe Chen, Zhuo Chen, Ziyi Chen, Jian Cong, Lelai Deng, Chuang Ding, Lu~Gao, et~al.,
\newblock ``Seed-tts: A family of high-quality versatile speech generation models,''
\newblock {\em arXiv preprint arXiv:2406.02430}, 2024.

\bibitem{wang2025spark}
Xinsheng Wang, Mingqi Jiang, Ziyang Ma, Ziyu Zhang, Songxiang Liu, Linqin Li, Zheng Liang, Qixi Zheng, Rui Wang, Xiaoqin Feng, et~al.,
\newblock ``Spark-tts: An efficient llm-based text-to-speech model with single-stream decoupled speech tokens,''
\newblock {\em arXiv preprint arXiv:2503.01710}, 2025.

\bibitem{wang2023cam++}
Hui Wang, Siqi Zheng, Yafeng Chen, Luyao Cheng, and Qian Chen,
\newblock ``Cam++: A fast and efficient network for speaker verification using context-aware masking,''
\newblock {\em arXiv preprint arXiv:2303.00332}, 2023.

\bibitem{desplanques2020ecapa}
Brecht Desplanques, Jenthe Thienpondt, and Kris Demuynck,
\newblock ``Ecapa-tdnn: Emphasized channel attention, propagation and aggregation in tdnn based speaker verification,''
\newblock {\em arXiv preprint arXiv:2005.07143}, 2020.

\bibitem{he2016deep}
Kaiming He, Xiangyu Zhang, Shaoqing Ren, and Jian Sun,
\newblock ``Deep residual learning for image recognition,''
\newblock in {\em Proceedings of the IEEE conference on computer vision and pattern recognition}, 2016, pp. 770--778.

\bibitem{wang2024advancing}
Shuai Wang, Zhengyang Chen, Bing Han, Hongji Wang, Chengdong Liang, Binbin Zhang, Xu~Xiang, Wen Ding, Johan Rohdin, Anna Silnova, et~al.,
\newblock ``Advancing speaker embedding learning: Wespeaker toolkit for research and production,''
\newblock {\em Speech Communication}, vol. 162, pp. 103104, 2024.

\bibitem{wang2023wespeaker}
Hongji Wang, Chengdong Liang, Shuai Wang, Zhengyang Chen, Binbin Zhang, Xu~Xiang, Yanlei Deng, and Yanmin Qian,
\newblock ``Wespeaker: A research and production oriented speaker embedding learning toolkit,''
\newblock in {\em IEEE International Conference on Acoustics, Speech and Signal Processing (ICASSP)}. IEEE, 2023, pp. 1--5.

\end{thebibliography}

\end{document}